\documentclass[12pt,twoside]{article}
\usepackage[mathscr]{eucal}
\usepackage{amsmath,amsfonts,amssymb,amsthm,mathabx,empheq}
\usepackage{times}
\usepackage{pdfsync}
\usepackage{cite}
\usepackage{url}
\usepackage{hyperref}
\usepackage{tensor}
\usepackage{color}
\usepackage{multicol}
\usepackage{bbold}

\advance\textheight\topskip
\allowdisplaybreaks[1]

\newcommand\td{\text{d}}

\newcommand{\p}{\partial}

\newcommand{\be}{\begin{equation}}
\newcommand{\ee}{\end{equation}}
\newcommand{\bea}{\begin{eqnarray}}
\newcommand{\eea}{\end{eqnarray}}
\def\nn{\nonumber}
\def\bz{\bar z}

\def\half{\frac12}

\def\bY{\bar Y}

\def\n{\nabla}
\def\bm{\bar{m}}

\def\bL{\bar{L}}

\def\bomega{\bar\omega}
\def\bY{\bar Y}

\def \sd {\delta\hspace{-0.50em}\slash\hspace{-0.05em}}

\newcommand*\xbar[1]{%
  \hbox{%
    \vbox{%
      \hrule height 0.5pt 
      \kern0.3ex
      \hbox{%
        \kern-0.0em
        \ensuremath{#1}%
        \kern-0.0em
      }%
    }%
  }%
}

\DeclareFontFamily{OT1}{rsfs}{} \DeclareFontShape{OT1}{rsfs}{m}{n}{
<-7> rsfs5 <7-10> rsfs7 <10-> rsfs10}{}
\DeclareMathAlphabet{\mycal}{OT1}{rsfs}{m}{n}

\begin{document}
\title{Near horizon gravitational charges}

\author{Hai-Shan Liu and Pujian Mao}

\date{}

\def\mytitle{Near horizon gravitational charges}

\addtolength{\headsep}{4pt}

\begin{centering}

  \vspace{1cm}

  \textbf{\Large{\mytitle}}

  \vspace{1.5cm}

  {\large Hai-Shan Liu and Pujian Mao}

\vspace{.5cm}

\vspace{.5cm}
\begin{minipage}{.9\textwidth}\small \it  \begin{center}
     Center for Joint Quantum Studies and Department of Physics,\\
     School of Science, Tianjin University, 135 Yaguan Road, Tianjin 300350, China
 \end{center}
\end{minipage}

\end{centering}


\vspace{1cm}

\begin{center}
\begin{minipage}{.9\textwidth}
  \textsc{Abstract}. In this paper, we study the near horizon symmetry and gravitational charges in the Newman-Penrose formalism. In particular we investigate the effect from topological terms. We find that the Pontryagin term and Gauss-Bonnet term have significant influence on the near horizon charges and bring interesting novel features. We show that the gravitational charge derived from a general class of topological terms including the Pontryagin term and Gauss-Bonnet term can be obtained from the ambiguities of the symplectic potential.
 \end{minipage}
\end{center}
\thispagestyle{empty}


\section{Introduction}

\paragraph{Motivation.} In gravitational theory, one usually deals with surface charges. Evaluating surface charge requires that the spacetime must be equipped with a boundary, i.e., a special hypersurface where to compute the surface charge. Two  natural choices for the boundary are usually applied which are the infinity and the horizon. At the spatial infinity, Arnowitt, Deser and Misner constructed, for the first time,  appropriate surface integrals for gravitational energy-momentum \cite{Arnowitt:1962hi}. Since causal line can not reach the spatial infinity, the ADM energy measures the total energy contained in the spacetime. Regrading to the system with gravitational waves that carry energy off, the Bondi energy was defined at null infinity \cite{Bondi:1962px}. The Bondi energy  can never increase. It is conserved if and only if there is no gravitational wave.

Horizons are one of the most fascinating objects in spacetime with Lorentzian signaturere. They are responsible for many remarkable semiclassical properties, such as the Hawking radiation \cite{Hawking:1975vcx} and the Bekenstein-Hawking entropy \cite{Bekenstein:1973ur,Bardeen:1973gs}. Based on the covariant phase space method \cite{Lee:1990nz,Iyer:1994ys}, Wald showed that Black hole entropy is just a surface charge evaluated on the Killing horizon  \cite{Wald:1993nt}. The first law of black hole mechanics is nothing but conservation of the surface charge between the horizon and infinity.\footnote{See also \cite{Hajian:2015xlp} for a generalization.}

Once spacetime is equipped with a boundary, one can on longer consider boundary term in the action as something trivial to the theory. The most significant example is the Gibbons-Hawking-York term \cite{York:1972sj,Gibbons:1976ue} that added to Einstein-Hilbert action to enforce Dirichlet's boundary conditions. Meanwhile boundary terms have their own  contributions to the surface charges \cite{Aros:1999id,Miskovic:2009bm,Araneda:2016iiy,Harlow:2019yfa,Freidel:2020xyx,Freidel:2020svx,Freidel:2020ayo,Freidel:2021fxf,Freidel:2021cjp,Godazgar:2020gqd,Godazgar:2020kqd,Ciambelli:2020qny}. However it is shown that the surface charge from the Pontryagin term and Gauss-Bonnet term at null infinity is absent at the leading and subleading order in the inverse of a large $r$ expansion \cite{Godazgar:2020gqd,Godazgar:2022foc}. This is somewhat under expected from the fact that those two are higher derivative terms which should have a faster fall-off behavior at large distance (see also \cite{Hou:2021bxz} for relevant evidence).

In this paper, we study the surface charge on the horizon with special emphasis on the effect from the Pontryagin term and Gauss-Bonnet term in four spacetmie dimensions. We derive the surface charge in the Newman-Penrose (NP) formalism \cite{Newman:1961qr}. The reason to work in first order formalism is that it is in particular efficient for computing the near horizon charges. Because the charge expression for first order theory will not involve any derivative on the fields and variation of the fields \cite{Julia:1998ys,Julia:2000er,Ashtekar:2008jw,Corichi:2013zza,Jacobson:2015uqa,Barnich:2016rwk,DePaoli:2018erh,Barnich:2019vzx,Oliveri:2019gvm,Barnich:2020ciy}. Since any field at most starts at $O(1)$ in the near horizon expansion, it is enough to just consider one order of the field from the solution space for computing the charge. In the NP formalism, the most important variable is the spin connection. Each component has very definite geometric meaning. Given the fact that the charge does not include any derivative on the fields and the near horizon charges have clear thermodynamic interpretation \cite{Wald:1993nt,Jacobson:1995ab,Adami:2021kvx}, the relation between geometry and thermodynamics becomes transparent when computing the near horizon charges in the NP formalism.

\paragraph{Main result and plan of the paper.} In the next Section, we derive the near horizon solution space for the NP equations with suitable gauge and boundary conditions. We compare our results with the ones in \cite{Donnay:2015abr} and \cite{Adami:2021nnf}. Section \ref{symmetry} is devoted to the near horizon symmetry and the transformation law of the near horizon fields. In the NP formalism, a gauge transformation is a combination of a diffeomorphism and a local Lorentz transformation. We show that with the required gauge and boundary conditions, the residual Lorentz transformation is completely fixed by the residual diffeomorphism. The near horizon symmetry consists of the near horizon supertranslations and superrotations. Our main results are presented in Section \ref{charge} where the surface charges are obtained. We first compute the surface charge from the Palatini action. Our result is consistent with \cite{Donnay:2015abr,Adami:2021nnf,Mao:2016pwq}. Then we compute the surface charge from the Holst term. We find that the Holst charge is a total derivative term. Consequently, the dual mass \cite{Godazgar:2018qpq} as the zero mode of the supertranslation charge is absent. Our next result is the charge from the Pontryagin term. We first show that the symplectic potential derived from a general class of terms that do not affect equations of motion including the Pontryagin term and Gauss-Bonnet term can be written as a combination of a $Y$ term and a $W$ term which are ambiguities of the symplectic structure. We then compute the near horizon Pontryagin charge. The last case of our investigation is the Gauss-Bonnet charge. The near horizon charges from the Pontryagin term and Gauss-Bonnet term start at the leading order in the near horizon expansion. A surprising feature of the Gauss-Bonnet charge is that it includes field which is not horizon data. The charges from the Palatini action and the Pontryagin term only involve the area of the horizon, the expansion of the horizon generator, and the rotation of reference frame on the horizon. While the Gauss-Bonnet charge involves the expansion of the other null direction, other than the horizon generator. In the last Section, we close with some final remarks. There are five appendices with some details of the computation in the main text.

\section{Near horizon solution space}
\label{solution}
Newman and Penrose \cite{Newman:1961qr} established a special tetrad formalism with four null basis vectors $e_1=l=e^2,\;e_2=n=e^1,\;e_3=m=-e^4,\;e_4=\bar{m}=-e^3$. The basis vectors $l$ and $n$ are real while $m$ and $\bm$ are complex conjugates of each other. The null basis vectors have the following orthogonality and normalization conditions
\be\label{tetradcondition}
l\cdot m=l\cdot\bm=n\cdot m=n\cdot\bm=0,\quad l\cdot n=1,\quad m\cdot\bm=-1.
\ee
The spacetime metric is obtained from the tetrad as
\be
g_{\mu\nu}=n_\mu l_\nu + l_\mu n_\nu - m_\mu {\bm}_\nu - m_\nu \bm_\mu.
\ee
The components of spin connection are labeled by twelve Greek symbols. Ten independent components of the Weyl tensor are represented by five complex scalars. The Ricci tensor is defined in terms of four real scalars and three complex scalars. We will focus on vacuum solution in this work. So all the components of Ricci tensor are zero. For the notations, we would refer to \cite{Chandrasekhar}.

In the NP formalism, by local Lorentz transformations, it is always possible to impose
\begin{align}
\pi=\kappa=\epsilon=0,\,\,\;\;\rho=\bar\rho,\;\;\,\,\tau=\bar\alpha+\beta.
\end{align}
According to the relations in Appendix \ref{relation}, such gauge choice means that $l$ is tangent to a null geodesic with affine parameter. The other null basis vectors are parallel transported along $l$. Moreover, $l$ is the gradient of a scalar field. It is of convenience to choose this scalar field as coordinate $u=x^1$ and take the affine parameter as coordinate $r=x^2$. For the rest two angular coordinates, we choose the stereographic coordinates $A=(z,\bz)$ which are related to the usual angular variables $(\theta,\phi)$ by $z=\cot\frac\theta2 e^{i\phi}$. The tetrad and the co-tetrad satisfying conditions in \eqref{tetradcondition} must have the forms
\begin{align}
&n=\frac{\p}{\p u} + U \frac{\p}{\p r} + X^A \frac{\p}{\p x^A},\nn\\
&l=\frac{\p}{\p r},\label{tetrad}\\
&m=\omega\frac{\p}{\p r} + L^A \frac{\p}{\p x^A},\nn
\end{align}
and
\begin{align}
&n=\left[-U-X^A(\xbar\omega L_A+\omega \bar L_A)\right]\td u + \td r + (\omega\bar L_A+\xbar\omega L_A) \td x^A,\nn\\
&l=\td u,\label{co-tetrad}\\
&m=-X^A L_A \td u + L_A \td x^A,\nn
\end{align}
where $L_AL^A=0,\;L_A\bar L^A=-1$. The line element is then
\begin{multline}
\td s^2 = -2 (U -\omega  \bomega ) \td u^2 + 2 \td u \td r \\
+g_{AB}\left[\td x^A + (L^A \bomega + \bL^A \omega -X^A)\td u \right]\left[\td x^B + (L^B \bomega + \bL^B \omega -X^B)\td u \right],
\end{multline}
where
\be
g_{AB}=-L_A \bL_B - \bL_A L_B.
\ee
The tetrad one form and the connection one form are defined as
\be
e^a=e^a_\mu \td x^\mu,\quad \Gamma_{ab}=\Gamma_{abc}e^c_\mu \td x^\mu .
\ee
The exact formulas of the form $e^a$ are given in \eqref{co-tetrad} and the connection one form $\Gamma_{ab}$ are given as
\bea
&& \Gamma_{12}=-(\gamma + \xbar\gamma) l + \xbar\tau m  +\tau \bm,\\
&&\Gamma_{13}= -\tau l + \rho m + \sigma \bm,\\
&&\Gamma_{23}=\xbar\nu l - \xbar\mu m - \xbar\lambda \bm ,\\
&&\Gamma_{34}=(\gamma -\xbar\gamma) l - (\alpha-\xbar\beta) m  + (\xbar\alpha- \beta ) \bm.
\eea

Boundary conditions are somehow tricky in the near horizon case. In principle, all fields can start at $O(1)$. Different choices will lead to the horizon with different property \cite{Grumiller:2019fmp}, such as Killing horizon, isolated horizon \cite{Ashtekar:2000sz,Ashtekar:2004cn},  non-expansion horizon, dynamical horizon, and so on so forth. The main boundary conditions that We will choose are
\be
U=O(r),\quad \omega=O(r),\quad \nu=O(r),\quad L^z=O(r),\quad X^A=O(r).
\ee
The first two conditions guarantee that the $r=0$ hypersurface is null and we take it as the horizon. The third condition leads to the fact that $n$ is the generator of the horizon. The forth one defines a conformally flat horizon metric. The last one means that $n=\frac{\p}{\p u}$ on the horizon. One can consider $u$ as the time direction on the horizon. Two additional conditions are
\be\label{add-condition1}
\text{Im} [\mu]=O(r),\quad \lambda=O(r).
\ee
Those conditions mean that the horizon generator $n$ is twist free and shear free. So $u$ is considered as a good time direction. The last condition is
\be\label{add-condition2}
\text{Im} [L^{\bz}]=O(r).
\ee
This condition can eliminate independent internal Lorentz transformation, which doesn't exist in metric theory, and thus leads the first order theory to be as close as the metric theory. At null infinity, the standard treatment is to fix the boundary metric to be a sphere. However this condition is too strong in the near horizon case which will eliminate near horizon superrotations.

The advantage of computing near horizon surface charge in first order formalism is the fact that the leading order charge only involves the leading order fields. So one does not really need to solve the radial NP equations as the leading order fields are integration constants which are free data.\footnote{An $r$-derivative may appear in the variation of fields along symmetry direction. One just needs to trace the precise radial equations for the involved fields. Nevertheless this will not happen for a fixed boundary case as we will show in the next section.} For completeness, we work out one more order solutions of the radial equations which would be useful for computing the subleading near horizon charges. The full vacuum NP equations are listed in Appendix \ref{NP}. The details of solving the NP equations are given in Appendix \ref{NU solution}. The solutions in near horizon expansion are
\bea
&&\Psi_0=\Psi_0^0  + \Psi_0^1 r + O(r^2),\label{psi0}\\
&&\rho=\rho_0 + (\rho_0^2 + \sigma_0 \xbar\sigma_0) r +   O(r^2),\label{rho}\\
&&\sigma=\sigma_0 + (2\rho_0 \sigma_0 + \Psi_0^0) r +  O(r^2),\label{sigma}\\
&&L^z= P \sigma_0 r +  \frac12 P (4\rho_0\sigma_0 + \Psi_0^0) r^2  + O(r^3),\label{Lz}\\
&&L^{\bz}=P + P\rho_0 r +  P (\rho_0^2 + \sigma_0\xbar\sigma_0) r^2  + O(r^3),\label{Lzb}\\
&&L_z=-\frac{1}{P} + \frac{\rho_0} {P} r + O(r^3),\\
&&L_{\bz}=\frac{\sigma_0}{P}r + \frac{\Psi_0^0}{P} r^2  + O(r^3),\\
&&\alpha=\alpha_0 + (\alpha_0 \rho_0 + \beta_0 \xbar\sigma_0) r + O(r^2),\label{alpha}\\
&&\beta=\beta_0 + (\alpha_0 \sigma_0+\beta_0\rho_0 + \Psi_1^0) r + O(r^2),\label{beta}\\
&&\omega=-\tau_0 r + O(r^2),\label{omega}\\
&&\Psi_1=\Psi_1^0 + (4\rho_0 \Psi_1^0 + \xbar\eth \Psi_0^0) r  + O(r^2),\label{psi1}\\
&& X^z=P \tau_0 r + O(r^2),\label{XA}\\
&&\mu=\mu_0 + (\mu_0\rho_0 + \Psi_2^0) r  + O(r^2),\label{mu}\\
&&\lambda=\mu_0\xbar\sigma_0 r + O(r^2), \label{lambda}\\
&&\Psi_2=\Psi_2^0 + (3\rho_0\Psi_2^0 + \xbar\eth\Psi_1^0) r + O(r^2),\label{psi2}\\
&&\gamma=\gamma_0 + (\alpha_0\xbar\alpha_0 + \beta_0\xbar\beta_0 + 2\alpha_0\beta_0 + \Psi_2^0) r + O(r^2),\label{gamma}\\
&& U=-(\gamma_0+\xbar\gamma_0) r  + O(r^2),\label{U}\\
&&\Psi_3=\Psi_3^0 + (2\rho_0\Psi_3^0 + \xbar\eth \Psi_2^0) r + O(r^2),\label{psi3}\\
&&\nu=(\alpha_0\mu_0 + \xbar\beta_0 \mu_0 + \Psi_3^0)r+O(r^2),\label{nu}\\
&&\Psi_4=\Psi_4^0 + (\rho_0\Psi_4^0 + \xbar\eth\Psi_3^0)r +O(r^2)\label{psi4},
\eea
where quantities with subscript $0$ are integration constants in $r$. The ``$\eth$'' operator is defined by
\begin{equation}\begin{split}
&\eth \eta^s=P\p_{\bz} \eta^s + 2 s\xbar\alpha^0 \eta^s,\\
&\xbar\eth \eta^s=P\p_z \eta^s -2 s \alpha^0 \eta^s,\nn
\end{split}\end{equation}
where $s$ is the spin weight of the field $\eta$. The spin weights of relevant fields are listed in Table \ref{t1}.
\begin{table}[ht]
\caption{Spin weights}\label{t1}
\begin{center}\begin{tabular}{|c|c|c|c|c|c|c|c|c|c|c|c|c|c|c|c|c|c}
\hline
& $\eth$ & $\p_u$ & $\gamma^0$ & $\nu^0$ & $\mu^0$ & $\sigma^0$ & $\lambda^0$  & $\Psi^0_4$ &  $\Psi^0_3$ & $\Psi^0_2$ & $\Psi^0_1$ & $\Psi_0^0$   \\
\hline
s & $1$& $0$& $0$& $-1$& $0$& $2$& $-2$  &
  $-2 $&  $-1$ & $0$ & $1$ & $2$    \\
\hline
\end{tabular}\end{center}\end{table}

The integration constants have the following constraints from the non-radial NP equations\footnote{There is one other possibility that $\mu_0=0$ and $\gamma_0$ is an arbitrary function.}
\bea
&&\alpha_0=\frac12 (\xbar\tau_0 + \p_z P),\label{alpha0}\\
&&\beta_0= \frac12(\tau_0  - \p_{\bz} P) ,\label{beta0}\\
&&\mu_0=-\p_u \ln P,\label{mu0}\\
&&\gamma_0=-\frac12 \mu_0 - \frac{\p_u \mu_0}{2\mu_0},\label{gamma0}\\
&&\Psi_4^0=0,\label{psi40}\\
&&\Psi_3^0=\xbar\eth\gamma_0 - \p_u \alpha_0 - \alpha_0 (\gamma_0-\xbar\gamma_0 + \mu_0),\label{psi30}\\
&&\Psi_2^0=(\gamma_0\xbar\gamma_0)\rho_0 - 2\alpha_0 \beta_0 - 2\alpha_0 \xbar\sigma_0 - \mu_0\rho_0 + P \p_z\xbar\alpha_0+P\p_z \beta_0 + \p_u\rho_0,\label{psi20}\\
&&\Psi_1^0=\xbar\alpha_0\rho_0 + \alpha_0\sigma_0 + \beta_0\rho_0 - \eth\rho_0 + \xbar\eth\sigma_0,\label{psi10}\\
&& \p_u \tau_0 =-\p_u\p_{\bz} P - 2\beta_0\mu_0 - 4\xbar \alpha_0 \mu_0 + 2\eth\gamma_0 - 2\eth\mu_0, \label{tau0}\\
&&\p_u \rho_0 = \rho_0(\gamma_0+\xbar\gamma_0) - 2\mu_0\rho_0 + P\p_{\bz} \alpha_0 + P \p_z \xbar\alpha_0 - 3\alpha_0 \xbar\alpha_0 - \beta_0 \xbar\beta_0, \label{rho0}\\
&&\p_u\sigma_0= -2\xbar\alpha_0 \beta_0 - 2 \beta_0^2 + 2\gamma_0\sigma_0 - \mu_0\sigma_0 + P\p_{\bz} \xbar\alpha_0 + P\p_{\bz} \beta_0. \label{sigma0}\\
&&\p_u \Psi_0^0 - 4\gamma_0 \Psi_0^0 + \mu_0 \Psi_0^0 = \eth \Psi_1^0 - 6 \beta_0 \Psi_1^0 - 6\xbar\alpha_0 \Psi_1^0 + 3\sigma_0 \Psi_2^0 \label{psi00}.
\eea
There is no constraint on the time evolution of $P$ which can characterize the flux going through the horizon. The horizon data are $\mu_0$ the expansion, $\gamma_0$ the surface gravity and $\tau_0$ the rotation of reference frame on the horizon. The structure of the solution space is quite different from the case at null infinity where it is normally to consider the Weyl tensor as free data, then the spin connection is determined by the Weyl tensor. But on the horizon the Bianchi identities can not lead to a peeling-off property. We do the inverse that the Weyl tensor is determined by the spin connection. Then the time evolution equations from the Bianchi identities are satisfied automatically. The practical reason of such treatment is that we care more about the local geometric structure on the horizon which is encoded in the spin connection.

The solution space we obtained is larger than that in \cite{Donnay:2015abr} where the authors considered a non-expanding horizon, i.e., $\mu_0=0$. While our solution space is smaller than that in \cite{Adami:2021nnf} where the authors consider the case with the shear of the horizon generator which can also characterize flux going through the horizon. To recover our solution space, one just needs to set the fields in \cite{Adami:2021nnf} to\footnote{Such choice of parameters is not included in the discussion in \cite{Adami:2021nnf}. However, as local functions, they are solution of Einstein equation.}
\be
{\cal U}^A\rightarrow0,\quad \eta\rightarrow 1,\quad  \Omega\rightarrow\frac{1}{P^2},\quad \gamma_{z\bz}\rightarrow-1,\quad \gamma_{zz}\rightarrow0.
\ee
and
\be
\kappa\rightarrow- 2\gamma_0,\quad  \Upsilon^z\rightarrow \frac{2\tau_0}{P}.
\ee
Other simplifications of \cite{Adami:2021nnf} in this case are
\be
{\cal D}_v=\p_v,\quad \Theta_l=\p_v\ln \Omega\rightarrow2\mu_0,\quad N^{AB}=0,\quad \Gamma=-2\kappa+\p_v\ln\Omega\rightarrow4\gamma_0 + 2\mu_0.
\ee
Those relations will be very useful later to compare our charge with the one in  \cite{Adami:2021nnf}.

\section{Near horizon symmetries}
\label{symmetry}

In the NP formalism, the gauge transformation of the tetrad and the spin connection is a combination of a diffeomorphism and a local Lorentz transformation. The transformation is given by
\be
\begin{split}
&\delta_{\xi,\omega}{e_a}^\mu ={\xi}^\nu\partial_\nu
{e_a}^\mu-\p_\nu{\xi}^\mu{e_a}^\nu +{\omega_a}^b{e_b}^\mu, \\
&\delta_{\xi, \omega} \Gamma_{a b c} = {\xi}^\nu \partial_\nu \Gamma_{a b c} - e_c^\mu \p_\mu {\omega}_{a b} + {\omega_a}^{d}\Gamma_{dbc}+ {\omega_b}^{d}\Gamma_{adc} + {\omega_c}^{d}\Gamma_{abd} .
\end{split}
\ee
The residual gauge transformation that preserved the gauge and boundary conditions is worked out in Appendix \ref{asg}. The symmetry parameters are given by $f(u,z,\bz)$ and $Y(z),\bY(\bz)$ which generate near horizon supertranslatoins and superrotations respectively. The associated residual gauge transformations are explicitly determined by the symmetry parameters as
\be\begin{split}
&\xi^u=f,\quad \xi^A=Y^A + \p_B f
  \int^r_0 \td r[L^A \bL^B + \bL^AL^B ],\\\nn
&\xi^r=-\p_u f r + \p_A
  f \int^r_0 \td r[\omega \bL^A + \bomega L^A - X^A],
\end{split}
\ee
and
\be\begin{split}\nn
&\omega^{12}=\p_u f + X^A \p_A f,\quad \omega^{13}= - \p_A f \int^r_0
\td r[\lambda L^A + \mu \bL^A],\\
&\omega^{23}= \bL^A \p_A f,\quad \omega^{34}=\frac12(\p_{\bz} \bY - \p_z Y ) + \p_A f \int^r_0
  \td r[(\bar{\alpha}-\beta) \bL^A - (\alpha - \bar{\beta}) L^A].
\end{split}
\ee
The constant order in $\xi^r$ is set to be zero by hand. The reason is that we want to fix the $r=0$ null hypersurface to be the boundary. Somehow this can be understood as the fact that the existence of a boundary at $r=0$ breaks the translational invariance along $r$ direction \cite{Donnay:2015abr,Adami:2021nnf}.

Acting the residual gauge transformation on the near horizon fields yields their transformation law as
\begin{align}
&\delta_{\xi,\omega} \frac{1}{P} = f\p_u \frac{1}{P}  +Y^A \p_A \frac{1}{P}  +  \frac12 \p_A Y^A  \frac{1}{P},\label{dP}\\
&\delta_{\xi,\omega} \p_z \ln P = f\p_u \p_z \ln P  + Y^A \p_A\p_z  \ln P  - \mu_0 \p_z f + \p_z Y \p_z \ln P -\frac12 \p_z^2Y,\label{dlnP}\\
&\delta_{\xi,\omega} \mu_0 =f\p_u \mu_0  + Y^A \p_A \mu_0  + \p_u f \mu_0, \label{dmu}\\
&\delta_{\xi,\omega} \frac{\mu_0}{P} =f\p_u \frac{\mu_0}{P}  + Y^A \p_A \frac{\mu_0}{P}  + \p_u f \frac{\mu_0}{P} +  \frac12 \p_A Y^A  \frac{\mu_0}{P}, \label{dmuP}\\
&\delta_{\xi,\omega} \gamma_0 = f\p_u \gamma_0 + Y^A \p_A \gamma_0  + \p_u f \gamma_0  - \frac12 \p_u^2 f,\label{dgamma}\\
&\delta_{\xi,\omega} \tau_0= f\p_u \tau_0 + Y^A \p_A \tau_0 - \frac12(\p_z Y - \p_{\bz} \bY )\tau_0 + 2 P\p_{\bz} f  \gamma_0 -  \p_u(P \p_{\bz} f). \label{dalphabeta}\\
&\delta_{\xi,\omega} \frac{\tau_0}{P}= f\p_u \frac{\tau_0}{P} + Y^A \p_A \frac{\tau_0}{P} + \p_{\bz} \bY \frac{\tau_0}{P} +  \p_{\bz} f  (2\gamma_0 + \mu_0 ) -  \p_u \p_{\bz} f. \label{dalphabetaP}\\
&\delta_{\xi,\omega} \rho_0 = f\p_u \rho_0  + Y^A \p_A \rho_0  - \p_u f \rho_0
 + P \p_{\bz}f \xbar\tau_0 + P\p_z f \tau_0 - P^2\p_z\p_{\bz} f.\label{drho}\\
 &\delta_{\xi,\omega} \frac{\rho_0}{P} = f\p_u \frac{\rho_0}{P}  + Y^A \p_A \frac{\rho_0}{P}  - \p_u f \frac{\rho_0}{P} + \frac12 \p_A Y^A \frac{\rho_0}{P}
 + \p_{\bz}f \xbar\tau_0 + \p_z f \tau_0 - P\p_z\p_{\bz} f.\label{drhoP}\\
&\delta_{\xi,\omega} \sigma_0 = f\p_u \sigma_0  + Y^A \p_A \sigma_0 - (\p_z Y -\p_{\bz} \bY) \sigma_0 - \p_u f \sigma_0\nn\\
&\hspace{7.5cm}  + 2P \p_{\bz}f \tau_0 -2 P\p_{\bz} f \p_{\bz} P - P^2\p_{\bz}^2 f.\label{dsigma}
\end{align}

For field dependent symmetry generators, one can define the adjusted Lie bracket that subtracts the changes in the symmetry transformation due to the variation of the fields \cite{Barnich:2010eb,Barnich:2011mi,Compere:2015knw}.  In the NP formalism, the adjusted bracket is defined as \cite{Barnich:2019vzx}
\begin{equation}
  \begin{split}
& [\delta_{\xi_1,\omega_1},\delta_{\xi_2,\omega_2}]\phi^\alpha=
-\delta_{\hat\xi,\hat\omega}\phi^\alpha,\\
& \hat\xi^\mu=[\xi_1,\xi_2]^\mu
-\delta_{\xi_1,\omega_1}{\xi}^\mu_2+\delta_{\xi_2,\omega_2}{\xi}^\mu_1,
  \\
 & {{(\hat{\omega})}_a}^b
 ={\xi_1}^\rho\p_\rho{\omega_{2a}}^b+{\omega_{1a}}^c{\omega_{2c}}^b
 -\delta_{\xi_1,\omega_1}{\omega_{2a}}^b
  -(1\leftrightarrow
  2),
\end{split}
\end{equation}
where $\phi^\alpha$ denotes an arbitrary field. Using this adjusted bracket, the near horizon symmetry parameters $(\xi[f,Y^A],\omega[f,Y^A])$
realize a symmetry algebra anywhere in the near horizon region,
\begin{equation}\label{NHSG}
  \begin{split}
  &\hat \xi=\xi[\hat f,\hat Y^A],\quad \hat
  \omega=\omega[\hat f,\hat Y^A],\\
  &\hat f=Y_1^A\p_A f_{2}+ f_{1}\p_u f_2 -(1\leftrightarrow
  2),\\
  & \hat Y^A=Y_1^B\p_B Y^A_2-Y_2^B\p_B Y^A_1.
\end{split}
\end{equation}
In particular, on the horizon (at the leading order of $r$), the symmetries form an algebra with the standard Lie bracket when $f$ and $Y^A$ are field independent. The symmetry algebra \eqref{NHSG} has several interesting subalgebra. Setting $f=T(z,\bz)+\frac{u}{2} D_A Y^A$ where $D_A$ is the covariant derivative associated to the horizon metric, one can recover the $\mathfrak{bms}_4$ algebra \cite{Barnich:2010eb}. By turning off $Y^A$, the generator $f$ is the T-witt generator in \cite{Adami:2020amw}. If we turn off $f$, the generators $Y^A$ form the 2d conformal symmetry algebra, i.e., two commuting copies of the Virasoro
algebra.

\section{Near horizon charges}
\label{charge}
In this section, we will compute the surface charge defined in \cite{Godazgar:2020kqd} for the Palatini action, Holst term, Pontryagin term, and Gauss-Bonnet term.

\subsection{Palatini action}
The Palatini Lagrangian is
\be\label{palatiniaction}
L_{Pa}=\frac{1}{32\pi G}\epsilon_{abcd} R^{ab}\wedge e^c \wedge e^d ,
\ee
where $R^{ab}=d \Gamma^{ab} + \Gamma^{ac}\wedge {\Gamma_c}^b$ is the curvature tensor. The surface charge from this Lagrangian is defined by\footnote{There is minus sign missing for the Lorentz charge in \cite{Godazgar:2020kqd}.}
\be\label{palatinicharge}
\sd {\cal H}_{Pa}= \frac{1}{16\pi G} \epsilon_{abcd} \int_{\partial \Sigma} (i_\xi \Gamma^{ab}  \delta e^c \wedge e^d + i_\xi e^c \delta \Gamma^{ab}\wedge e^d - \omega^{ab} \delta e^c\wedge e^d),
\ee
where $\partial \Sigma$ can be any constant-$u$ two surface on the horizon to evaluate the surface charge. The symmetry parameters should be field independent which was assumed to obtain this expression in \cite{Godazgar:2020kqd}. Inserting the solutions and the symmetry parameters into the charge gives
\begin{multline}\label{palatini}
\sd {\cal H}_{Pa}=\frac{1}{8\pi G}\int_{\partial \Sigma} \td z \td \bz \delta\left[\frac{1}{P^2}\left(\p_u f - 2f\mu_0 - Y\frac{\xbar\tau_0}{P}  - \bY\frac{\tau_0}{P} \right )\right]\\
+ \frac{1}{8\pi G}\int_{\partial \Sigma} \td z \td \bz f \left(\mu_0 -2 \gamma_0\right) \delta \frac{1}{P^2}.
\end{multline}
This charge matches the one in \cite{Adami:2021nnf} explicitly using the relations by the end of Section \ref{solution}. Note that fixing $\eta=1$ requires that $W=2\p_v T$ in \cite{Adami:2021nnf} which also matches our near horizon symmetry generators in Section \ref{symmetry}.

In general, the surface charge of a theory with propagating degrees of freedom is not integrable. One can use the Barnich-Troessaert prescription \cite{Barnich:2011mi} to split the charge into an integrable part and a flux part, $\sd {\cal H}_{(\xi,\omega)} =\delta{\cal H}^I_{(\xi,\omega)}  + {\cal F}_{(\xi,\omega)} (\delta \phi^\alpha;\phi^\alpha)$, such that the surface charges satisfy the modified bracket
\be\label{bracket}
\begin{split}
&\delta_{(\xi_2,\omega_2)} {\cal H}^I_{(\xi_1,\omega_1)} + {\cal F}_{(\xi_2,\omega_2)} (\delta_{(\xi_1,\omega_1)} \phi^\alpha;\phi^\alpha):=\{ {\cal H}^I_{(\xi_1,\omega_1)},{\cal H}^I_{(\xi_2,\omega_2)} \}_{MB} \\
&\{ {\cal H}^I_{(\xi_1,\omega_1)},{\cal H}^I_{(\xi_2,\omega_2)} \}_{MB}={\cal H}^I_{(\hat{\xi},\hat{\omega})} + K_{(\xi_1,\omega_1),(\xi_2,\omega_2)},
\end{split}
\ee
where $(\hat{\xi},\hat{\omega})$ are defined from the adjusted bracket of the symmetry parameters and $K_{(\xi_1,\omega_1),(\xi_2,\omega_2)}$ is the 2-cocycle term. It is shown in \cite{Adami:2021nnf} that the Barnich-Troessaert prescription yields the integrable part of
\eqref{palatini} as
\be
{\cal H}_{Pa}^I=\frac{1}{8\pi G}\int_{\partial \Sigma} \td z \td \bz \frac{1}{P^2}\left(\p_u f - f\mu_0 - 2f \gamma_0 - Y\frac{\xbar\tau_0}{P}  - \bY\frac{\tau_0}{P} \right ),
\ee
while the flux part as
\be
{\cal F}_{Pa}=\frac{1}{8\pi G}\int_{\partial \Sigma} \td z \td \bz \frac{1}{P^2} f \delta (2 \gamma_0 - \mu_0).
\ee
There is no 2-cocycle term in this case. By virtue of the surface charge algebra, one can read the balance equation for the charges as
\be\label{balance}
\begin{split}
\frac{\p}{\p u} {\cal H}_{Pa}^I&=\delta_{\frac{\p}{\p u}} {\cal H}_{Pa}^I+{{\cal H}^I_{Pa}}_{(\p_u \xi, \hat\omega)}=-{{\cal F}_{Pa}}_{\frac{\p}{\p u}} (\delta_{(\xi,\omega)} \phi^\alpha;\phi^\alpha) + K_{(\xi,\omega),(\frac{\p}{\p u},\omega_2)}\\
&=\frac{1}{8\pi G}\int_{\partial \Sigma} \td z \td \bz \frac{1}{P^2}\left[ 2\p_u f\mu_0 + f \p_u( \mu_0 - 2 \gamma_0) + Y^A \p_A( \mu_0 - 2 \gamma_0)\right] .
\end{split}
\ee
Recalling the relation in \eqref{mu0} and \eqref{gamma0}, the non-conservation of the surface charge is induced by the flux going through the horizon which is characterized by the unconstrained $u$-dependence of $P$.

\subsection{Holst term}

The Holst term is
\be
L_H=\frac{it}{16\pi G} R_{ab}\wedge e^a \wedge e^b,
\ee
where $t$ is the Holst term parameter. The Holst term is not a boundary term, but the addition of this term does not affect the equations of motion from the Palatini action \cite{Godazgar:2020kqd,DePaoli:2018erh}. The contribution in the charge from the Holst term is given by \cite{Godazgar:2020kqd}
\be
\sd {\cal H}_H= \frac{it}{8\pi G}  \int_{\partial \Sigma} (i_\xi \Gamma^{ab}  \delta e_a \wedge e_b + i_\xi e_a \delta \Gamma^{ab}\wedge e_b - \omega^{ab} \delta e_a\wedge e_b).
\ee
When the near horizon solution and symmetries are inserted, the near horizon Holst charge reads
\be
\sd {\cal H}_H =\frac{i t}{16\pi G} \int_{\partial \Sigma} \td z \td \bz \delta\left[ \p_{z}\frac{ Y}{P^2}-  \p_{\bz}\frac{ \bY}{P^2} \right].
\ee
Defining ${\cal Y}=i\frac{ Y}{P^2}$ and  ${\cal \bar{Y}}=-i \frac{\bY}{P^2}$, the Holst charge can be written as a total derivative
\be
\sd {\cal H}_H =\frac{t}{16\pi G} \int_{\partial \Sigma} \td z \td \bz \delta\left[ \p_{A} {\cal Y}^A \right].
\ee
It may be a puzzling fact that the near horizon Holst charge is vanishing which is proposed to count for the dual charges \cite{Godazgar:2020gqd,Godazgar:2020kqd}. To understand this fact better, we check the near horizon charges for the Taub-NUT solution. The solution in the near horizon expansion is given in Appendix \ref{TN}. The near horizon Palatini charge for the Taub-NUT solution is
\begin{multline}
{\cal H}_{PaT}^I=\frac{1}{8\pi G}\int_{\partial \Sigma} \td z \td \bz \frac{4}{(1+z\bz)^2}\bigg[\p_u f(r_+^2 + N^2) - f  \sqrt{M^2+N^2} \\
+ Y\frac{2i N \sqrt{M^2+N^2} \bz}{(1+z\bz)}  - \bY\frac{2i N \sqrt{M^2+N^2} z}{(1+z\bz)} \bigg].
\end{multline}
Clearly the mass parameter $M$ and the NUT parameter $N$ contribute equally to the global charge, i.e., the zero mode of the supertranslation charge with $f=1,Y=0$ case. At null infinity, the NUT parameter has the interpretation of a dual mass \cite{Godazgar:2019dkh}.
It is worthwhile to point out that the addition of the Holst term does not enhance the near horizon symmetries.\footnote{In the linearized cases, a duality symmetric theory can have enhanced asymptotic symmetries and charges, see for instance \cite{Barnich:2008ts,Hosseinzadeh:2018dkh}.} So the charge derived from the Holst term is the dual part of the full Palatini-Holst charge. Then full global charge consists of both mass and NUT parameter. In this sense, we do not see any conceptual problem having a vanishing Holst charge. Note also that the structure of the near horizon solution space is very different from the one at null infinity as briefly discussed in previous section. One can also understand the absence of the dual charge from a different point of view. The dual charge should be located at the ``conjugate'' point of the
charge. When the charge derived from the Palatini action is inside the horizon, the
dual charge should be outside the horizon. Hence the near horizon dual charge vanishes.

\subsection{Pontryagin term}

The Pontryagin term is given by
\be\label{P}
L_{P}=\frac{1}{32\pi G}  R_{ab}\wedge R^{ab} .
\ee
This term can be written as a boundary term on the horizon
\be\label{Pb}
L_{Pb}=\frac{1}{32\pi G}\td \left(\Gamma^{ab}\wedge \td \Gamma_{ab} - \frac23 {\Gamma^a}_b \wedge {\Gamma^b}_c \wedge {\Gamma^c}_a \right) .
\ee
A boundary term can only contribute to the symplectic structure through the corner symplectic potential \cite{Harlow:2019yfa,Freidel:2020xyx}. The corner symplectic potential from \eqref{Pb} is
\be
\vartheta_P=\frac{1}{32\pi G}  \Gamma^{ab}\wedge \delta \Gamma_{ab}.
\ee
The surface charge obtained from this corner symplectic potential is
\be\label{Pcharge}
\sd {\cal H}_{P}= \frac{1}{16\pi G}\int_{\partial \Sigma}
\delta \Gamma^{ab}\wedge \delta_{\xi,\omega} \Gamma_{ab}
\ee

Alternatively in \cite{Godazgar:2020gqd}, a symplectic potential was derived directly from the variation of \eqref{P}, which is
\be\label{thetaP}
\theta_P=\frac{1}{16\pi G} \delta \Gamma_{ab}\wedge R^{ab}.
\ee
Eventually, the same surface charge \eqref{Pcharge} can be obtained. This symplectic potential can be rewritten as
\be\label{Y}
\theta_P = \frac{1}{32\pi G} \delta\left(\Gamma^{ab}\wedge \td \Gamma_{ab} - \frac23 {\Gamma^a}_b \wedge {\Gamma^b}_c \wedge {\Gamma^c}_a\right) + \frac{1}{32\pi G} \td ( \Gamma_{ab}\wedge \delta\Gamma^{ab}).
\ee
The two terms on the right hand side are a $W$ term and a $Y$ term in symplectic potential. Thus the Pontryagin charge in \cite{Godazgar:2020gqd} can be obtained from the ambiguities of the symplectic structure.

In the covariant phase space method, a boundary term can only lead to a $W$ term in the symplectic potential. It is very surprising that a $Y$ term arises from the Pontryagin term. Actually, the Lagrangian \eqref{P} is different from \eqref{Pb} with an extra trivial term
\be\label{surprising}
y_P=\frac{1}{32\pi G}  \Gamma^{ab}\wedge \td^2 \Gamma_{ab} .
\ee
In the variation of \eqref{P}, another trivial term\footnote{This term is the equation of motion from the variation of \eqref{P}.}
\be
\frac{1}{16\pi G}  \delta\Gamma^{ab}\wedge \td^2 \Gamma_{ab}
\ee
has been dropped. Including those two terms, the variation of \eqref{P} leads to
\be
\delta \left(L_P - y_P\right)=\frac{1}{32\pi G} \td \delta\left(\Gamma^{ab}\wedge \td \Gamma_{ab} - \frac23 {\Gamma^a}_b \wedge {\Gamma^b}_c \wedge {\Gamma^c}_a\right).
\ee
This will only lead to a $W$ term in the symplectic potential as it should be when considering \eqref{Pb}.

As demonstrated in \cite{Godazgar:2020gqd}, first order formalism is best suited to an analysis of surface charges, in particular for tracing the contribution from topological terms. However such idea was not implemented explicitly to the Pontryagin term. Here we fill in this gap and write the Pontryagin term in a complete first order form. We consider the curvature tensor as independent dynamical variable. The relation between the curvature tensor and the spin connection is a consequence of equation of motion. So the curvature tensor is an auxiliary field. The Lagrangian that is on-shell equivalent to the Pontryagin one is
\be\label{Pontryagin}
L_{PF}=\frac{1}{16\pi G} \left[ {\cal R}_{ab}\wedge \left(\td \Gamma^{ab} + \Gamma^{ac}\wedge {\Gamma_c}^b\right) - \frac12 {\cal R}_{ab}\wedge {\cal R}^{ab} \right].
\ee
The equations of motion derived from this Lagrangian are as follows. The variation on ${\cal R}_{ab}$ will lead to
\be\label{curvature}
{\cal R}^{ab}=d \Gamma^{ab} + \Gamma^{ac}\wedge {\Gamma_c}^b.
\ee
Hence ${\cal R}^{ab}$ is the curvature tensor. Then variation on $\Gamma^{ab}$ from this part will lead to the Bianchi identity of the curvature tensor once \eqref{curvature} applied. So this will not modify the equation of motion derived from the Palatini action when doing variation on $\Gamma^{ab}$. The Lagrangian \eqref{Pontryagin} is not a boundary term with the auxiliary field, the resulting $\theta$ term is
\be
\theta_{PF}=\frac{1}{16\pi G}\delta \Gamma^{ab}\wedge R_{ab},
\ee
which is the same as \eqref{thetaP}.

The way that we rewrite the Pontryagin term can be generalized to arbitrary fields. In the form language, one can always add the term
\be\label{Yaction}
L_A=2 A \wedge \td B -  A\wedge A,
\ee
to a Lagrangian without changing its on-shell solution where $A$ is a two form and $B$ is a one form. Such term can be modified by equation of motion into a boundary term in second order form $\td (B\wedge \td B)$. As highlighted in \cite{Godazgar:2020gqd}, one should include all that type of terms when finding asymptotic gravitational charges. The symplectic potential derived from \eqref{Yaction} is
\be
\theta_A =2 A \wedge \delta B.
\ee
On-shell this term becomes
\be
\theta_A =2 \td B \wedge \delta B.
\ee
Such term can be written as a combination of a $Y$ term and a $W$ term
\be
\theta_A =2 \td B \wedge \delta B= \td (B\wedge \delta B) +   \delta(\td B \wedge B).
\ee
Thus we have shown that the symplectic potential derived from a general glass of terms that do not affect the equations of motion for the theory of interest can be obtained from the ambiguities of the symplectic structure.

Inserting the near horizon solution and symmetry parameters from previous sections into \eqref{Pcharge}, the Pontryagin charge is obtained as
\begin{multline}\label{pontryagincharge}
\sd {\cal H}_{P}=\frac{1}{8\pi G}\int_{\partial \Sigma} \td z \td \bz \bigg[\delta_{\xi,\omega}\frac{\xbar\tau_0}{P}\delta\frac{ \tau_0}{P} - \delta\frac{\xbar\tau_0}{P}\delta_{\xi,\omega}\frac{\tau_0}{P}\\
+\delta\p_z\ln P \delta_{\xi,\omega} \p_{\bz} \ln P - \delta_{\xi,\omega}\p_z\ln P \delta \p_{\bz} \ln P\bigg],
\end{multline}
This charge is purely imaginary. So an imaginary unit $i$ should be included in the coupling constant for the Pontryagin term similar to the Holst case.
Including the Pontryagin part, the full charge algebra can be written as
\begin{multline}
\delta_{(\xi_2,\omega_2)} \left( {{\cal H}^I_{Pa}}_{(\xi_1,\omega_1)} + {{\cal H}^I_{P}}_{(\xi_1,\omega_1)}\right) + {{\cal F}_{Pa}}_{(\xi_2,\omega_2)} (\delta_{(\xi_1,\omega_1)} \phi^\alpha;\phi^\alpha) \\ + {{\cal F}_{P}}_{(\xi_2,\omega_2)} (\delta_{(\xi_1,\omega_1)} \phi^\alpha;\phi^\alpha)
={{\cal H}^I_{Pa}}_{(\hat{\xi},\hat{\omega})} + {{\cal H}^I_{P}}_{(\hat{\xi},\hat{\omega})} + K_{(\xi_1,\omega_1),(\xi_2,\omega_2)}.
\end{multline}
As shown previously, the charges from the Palatini action satisfy the modified charge bracket with no 2-cocycle term. According to the Barnich-Troessaert prescription, the Pontryagin charge \eqref{pontryagincharge} only contributes to the flux part of the full charge. This will lead to a 2-cocycle term
\be
{{\cal F}_{P}}_{(\xi_2,\omega_2)} (\delta_{(\xi_1,\omega_1)} \phi^\alpha;\phi^\alpha)
=  K_{P(\xi_1,\omega_1),(\xi_2,\omega_2)}.
\ee
Inserting \eqref{dlnP} and \eqref{dalphabetaP} into the 2-cocycle term, it becomes
\begin{multline}\label{KP}
{K_{P}}_{(\xi_1,\omega_1),(\xi_2,\omega_2)}=
\frac{1}{8\pi G}\int_{\partial \Sigma} \td z \td \bz \bigg[\bigg(f_2\p_u \p_{\bz} \ln P  + Y^A_2 \p_A\p_{\bz}  \ln P - \frac12 \p_{\bz}^2 \bY_2  - \mu_0 \p_{\bz} f_2\\
+ \p_{\bz} \bY_2 \p_{\bz} \ln P\bigg)
\times  \left( f_1\p_u \p_z \ln P  + Y^A_1 \p_A\p_z  \ln P - \frac12 \p_z^2 Y_1  - \mu_0 \p_z f_1 + \p_z Y_1 \p_z \ln P\right)\\  + \left(f_2\p_u \frac{\xbar\tau_0}{P} + Y^A_2 \p_A \frac{\xbar\tau_0}{P} + \p_z Y_2 \frac{\xbar\tau_0}{P} +  \p_z f_2  (2\gamma_0 + \mu_0 ) -  \p_u \p_z f_2\right)\\
\times \left(f_1\p_u \frac{\tau_0}{P} + Y^A_1 \p_A \frac{\tau_0}{P} + \p_{\bz} {\bY}_1 \frac{\tau_0}{P} +  \p_{\bz} f_1  (2\gamma_0 + \mu_0 ) -  \p_u \p_{\bz} f_1\right)
- (1 \leftrightarrow 2) \bigg].
\end{multline}
Direct computation shows that the 2-cocycle term \eqref{KP} satisfies the suitably generalized cocycle condition
\be
{K_{P}}_{[(\xi_1,\omega_1),(\xi_2,\omega_2)],(\xi_3,\omega_3)}-\delta_3 {K_{P}}_{(\xi_1,\omega_1),(\xi_2,\omega_2)} + \text{cyclic}(1,2,3)=0.
\ee
Note that our symmetry parameters $(f,Y,\bY)$ are field independent in the present case.
In principle, one can change the representative of the symmetry parameters along the lines of  \cite{Adami:2020ugu,Ruzziconi:2020wrb,Adami:2021sko} to simplify the expression of the 2-cocycle term. In particular, it is important to find the slicing in which the 2-cocycle term is a constant in the mode expansion, i.e., becoming the central extension term.  For instance, defining $\bar f=f\p_u \ln P  + Y^A \p_A \ln P - \frac12 \p_A Y^A$ and considering $\bar f$ as field independent will significantly simplify the first two lines of the 2-cocycle term. However the charge expression of Palatini action \eqref{palatinicharge} requires that the symmetry parameters $f$ and $Y^A$ are field independent \cite{Godazgar:2020kqd}. We can not change slicing in this expression. The symmetry parameters $Y^A$ in our case are (anti) holomorphic functions of the angular coordinates. Such properties must be kept to preserve the boundary conditions. Hence the freedom in choosing slicing for $Y^A$ is very limited. Those issues will be fixed elsewhere. At last, the balance equation \eqref{balance} will not be affected by the Pontryagin term as the flux term from the Pontryagin term and the 2-cocycle term are canceled.

It might be an unsatisfactory fact that the Pontryagin term does not give a integrable part of the charge. However this is a vary natural choice according to the expression of the Pontryagin charge \eqref{pontryagincharge}. And the 2-cocycle term may have more applications than the charge itself at the horizon, see, for instance, the relevance to black hole entry \cite{Strominger:1997eq,Carlip:1998wz,Carlip:2002be,Guica:2008mu,Majhi:2012tf,Carlip:2017xne,Carlip:2019dbu}. Nevertheless, one can always use the ambiguity in the separation of the charge to change the integrable and non-integrable piece, see, e.g., the discussion in \cite{Adami:2020amw}. For instance, one can write the Pontryagin charge with the integrable part as
\be
{\cal H}_{P  (\xi,\omega)}^I=\frac{1}{8\pi G}\int_{\partial \Sigma} \td z \td \bz \bigg[\frac{ \tau_0}{P}\delta_{\xi,\omega}\frac{\xbar\tau_0}{P} -  \frac{\xbar\tau_0}{P}\delta_{\xi,\omega}\frac{\tau_0}{P} +\p_z\ln P \delta_{\xi,\omega} \p_{\bz} \ln P - \p_{\bz} \ln P\delta_{\xi,\omega}\p_z\ln P  \bigg],
\ee
and the flux part as
\begin{multline}
{\cal F}_{P (\xi,\omega)} (\delta \phi^\alpha;\phi^\alpha)=\frac{1}{8\pi G}\int_{\partial \Sigma} \td z \td \bz \bigg[ \frac{\xbar\tau_0}{P}\delta(\delta_{\xi,\omega}\frac{\tau_0}{P}) - \frac{ \tau_0}{P}  \delta( \delta_{\xi,\omega}\frac{\xbar\tau_0}{P})\\
 + \p_{\bz} \ln P \delta(\delta_{\xi,\omega}\p_z\ln P ) - \p_z\ln P \delta(\delta_{\xi,\omega} \p_{\bz} \ln P)\bigg],
\end{multline}
where both the integrable part and the flux part are purely imaginary. Now the charge satisfy the modified bracket with a 2-cocycle term
\begin{multline}
K_{P(\xi_1,\omega_1),(\xi_2,\omega_2)}=\frac{1}{8\pi G}\int_{\partial \Sigma} \td z \td \bz \bigg[\delta_{\xi_1,\omega_1}\frac{\xbar\tau_0}{P}\delta_{\xi_2,\omega_2}\frac{ \tau_0}{P} - \delta_{\xi_2,\omega_2}\frac{\xbar\tau_0}{P}\delta_{\xi_1,\omega_1}\frac{\tau_0}{P}\\
+ \delta_{\xi_1,\omega_1} \p_{\bz} \ln P\delta_{\xi_2,\omega_2}\p_z\ln P -  \delta_{\xi_2,\omega_2} \p_{\bz} \ln P\delta_{\xi_1,\omega_1}\p_z\ln P\bigg].
\end{multline}
This 2-cocycle term satisfies the generalized cocycle condition. The balance equation \eqref{balance} in this case can be verified simply using the fact that $\frac{\p}{\p u}$ commutes with all other symmetries. One can check the Pontryagin charge for the Taub-NUT solution simply by taking advantage of \eqref{taubnutNP}. The Taub-NUT charge is
\begin{multline}
{\cal H}_{PT  (\xi,\omega)}^I=\frac{1}{8\pi G}\int_{\partial \Sigma} \td z \td \bz \bigg[
\p_{z} f \frac{i N  z}{2r_+^2(1+z\bz)} -  \p_u \p_{z} f \frac{i N z}{r_+(1+z\bz)}\\
+\frac{2 M}{r_+}\frac{\bz}{(1+z\bz)^2}\big(Y-z\p_z Y-\frac{r_+}{4 M}(1+z\bz)\p_{\bz}^2 \bar{Y}\big)
  - c.c \bigg].
\end{multline}
In particular, the zero mode of supertranslation $(f=1,Y^A=0)$ charge vanishes. The first mode of the supertranslation $(f=z\bz,Y^A=0)$ charge reads
\be\label{MP}
M_P=\frac{1}{8\pi G}\frac{i N}{r_+^2}\int_{\partial \Sigma} \td z \td \bz
 \frac{  z\bz}{(1+z\bz)}.
\ee
The zero mode of superrotation $(f=0,Y^z=Y^{\bz}=1)$ charges is
\be\label{JA}
J_P=\frac{1}{4\pi G}\frac{ M}{r_+} \int_{\partial \Sigma} \td z \td \bz \left[ \frac{\bz-z}{(1 + z \bz)^2} \right].
\ee
Note that the mass parameter $M$ and NUT parameter $N$, though being constant, are dynamical variables. Strictly speaking, the horizon of the Taub-NUT solution is not a fixed two-sphere because of its $M$ and $N$ dependence. To finally evaluate the surface charges on the horizon, one has to first fix the shape of the horizon which will reduce the near horizon symmetries. Moreover, the shape of the horizon can be deformed by residual gauge transformations, see, i.e., the angular dependent horizon of the supertranslated Schwarzschild black hole \cite{Compere:2016hzt,Hawking:2016sgy,Lin:2020gva}. Hence we leave the integral in \eqref{MP} and \eqref{JA} in the present form\footnote{When computed on a fixed sphere, the integrand in \eqref{MP} is divergent on the north pole $\theta=0$. The integral in \eqref{JA} vanishes. But the structure of the charge \eqref{JA} is of interest because of its connection to the Gauss-Bonnet charges as we will discuss in the next subsection.} and it should be just a constant factor.

\subsection{Gauss-Bonnet term}

The Gauss-Bonnet term is given by
\be\label{GB}
L_{GB}=\frac{1}{32\pi G}\epsilon_{abcd}  R^{ab}\wedge R^{cd} .
\ee
This term can be written as a boundary term as
\be\label{GBb}
L_{GBb}=\frac{1}{32\pi G}\epsilon_{abcd} \td \left(\Gamma^{ab}\wedge \td \Gamma^{cd} \right) .
\ee
The corner symplectic potential from \eqref{GBb} is
\be
\vartheta_{GB}=\frac{1}{32\pi G}\epsilon_{abcd} \Gamma^{ab}\wedge \delta \Gamma^{cd}.
\ee
The surface charge derived from this corner symplectic potential is
\be\label{GBcharge}
\sd {\cal H}_{GB}= \frac{1}{16\pi G}\epsilon_{abcd}\int_{\partial \Sigma}
\delta \Gamma^{ab}\wedge \delta_{\xi,\omega} \Gamma^{cd}.
\ee
This surface charge can be also obtained from the symplectic potential derived from the variation of  \eqref{GB}. The symplectic potential of \eqref{GB} is \cite{Godazgar:2020gqd}
\be
\theta_{GB}=\frac{1}{16\pi G}\epsilon_{abcd} \delta \Gamma^{ab}\wedge R^{cd},
\ee
which can be written as the sum of a $W$ term and a $Y$ term
\be
\theta_{GB} = \frac{1}{32\pi G}\epsilon_{abcd} \delta\left(\Gamma^{ab}\wedge \td \Gamma^{cd} \right) + \frac{1}{32\pi G} \epsilon_{abcd} \td ( \Gamma^{ab}\wedge \delta\Gamma^{cd}).
\ee
Eventually, the same surface charge \eqref{GBcharge} can be derived from this symplectic potential. Similar to the Pontryagin case, the Lagrangian \eqref{GB} can be written as a boundary term, i.e., \eqref{GBb}, plus an extra trivial term \be\label{surprisingGB}
y_{GB}=\frac{1}{32\pi G}\epsilon_{abcd}  \Gamma^{ab}\wedge \td^2 \Gamma^{cd} .
\ee
Including this term and keeping the trivial term from the equation of motion, the variation of \eqref{GB} leads to
\be
\delta \left(L_{GB} - y_{GB}\right)=\frac{1}{32\pi G}\epsilon_{abcd}\td \delta\left(\Gamma^{ab}\wedge \td \Gamma^{cd} \right),
\ee
as it should be when considering the Lagrangian \eqref{GBb}.

The surface charge \eqref{GBcharge} can also be deduced from a first order form as proposed in previous subsection. The first order term that leads to \eqref{GBcharge} is
\be\label{Gauss-Bonnet}
L_{GBF}=\frac{1}{16\pi G} \epsilon_{abcd}\left[ {\cal R}^{ab}\wedge  \td \Gamma^{cd} - \frac12 {\cal R}^{ab}\wedge {\cal R}^{cd} \right].
\ee
This term is not equivalent to the Gauss-Bonnet term in the sense that $ {\cal R}^{ab}$ is not the curvature tensor on-shell. However if the charge from the Gauss-Bonnet term has any interesting feature in four dimensions, one can not distinguish that charge is from  the Gauss-Bonnet term or from \eqref{Gauss-Bonnet}.

Inserting the solution and symmetry parameters into the surface charge \eqref{GBcharge}, we obtain
\begin{multline}
\sd {\cal H}_{GB}=\frac{1}{8\pi G}\int_{\partial \Sigma} \td z \td \bz \bigg[\delta\frac{\xbar\tau_0}{P}\delta_{\xi,\omega}\p_{\bz} \ln P +\delta \frac{\tau_0}{P} \delta_{\xi,\omega}\p_z \ln P
+2 \delta\frac{\rho_0}{P} \delta_{\xi,\omega} \frac{\mu_0}{P}\\
-2\delta \frac{\mu_0}{P}\delta_{\xi,\omega}\frac{\rho_0}{P} - \delta_{\xi,\omega}\frac{\xbar\tau_0}{P} \delta\p_{\bz} \ln P-\delta_{\xi,\omega} \frac{\tau_0}{P}  \delta\p_z \ln P\bigg].
\end{multline}
The Gauss-Bonnet charge is real. It is very surprising that $\rho_0$ term appears in the charge. The geometric meaning of $\rho_0$ is the expansion of $l$ which is another null direction other than the horizon generator $n$. So $\rho_0$ does not represent any horizon information. The interpretation of this term in a horizon charge is not clear and should be stressed elsewhere. Similar to the Pontryagin case, the Gauss-Bonnet charge is a flux part in the Barnich-Troessaert prescription which leads to a 2-cocycle term. Inserting \eqref{dlnP},  \eqref{dmuP}, \eqref{dalphabetaP} and \eqref{drhoP}, the 2-cocycle term is reduced to
\begin{multline}\label{KGB}
{K_{GB}}_{(\xi_1,\omega_1),(\xi_2,\omega_2)}= \frac{1}{8\pi G}\int_{\partial \Sigma} \td z \td \bz \bigg[\bigg(f_2\p_u \p_{\bz} \ln P  + Y^A_2 \p_A\p_{\bz}  \ln P + \frac12 \p_{\bz}^2 \bY_2 \frac{1}{P} - \mu_0 \p_{\bz} f_2\\
+ \p_{\bz} \bY_2 \p_{\bz} \ln P\bigg)
\times \left( f_1\p_u \frac{\xbar\tau_0}{P} + Y^A_1 \p_A \frac{\xbar\tau_0}{P} + \p_z Y_1 \frac{\xbar\tau_0}{P} +  \p_z f_1  (2\gamma_0 + \mu_0 ) -  \p_u \p_z f_1\right)\\
+\left( f_1\p_u \frac{\rho_0}{P}  + Y^A_1 \p_A \frac{\rho_0}{P}  - \p_u f_1 \frac{\rho_0}{P} + \frac12 \p_A Y^A_1 \frac{\rho_0}{P}
 + \p_{\bz}f_1 \xbar\tau_0 + \p_z f_1 \tau_0 - P\p_z\p_{\bz} f_1\right)\\
\times \left(f_2\p_u \frac{\mu_0}{P}  + Y^A_2 \p_A \frac{\mu_0}{P}  + \p_u f_2 \frac{\mu_0}{P}\right) + c.c - (1 \leftrightarrow 2)\bigg].
\end{multline}
For field independent symmetry parameters $(f,Y,\bY)$, we prove that the 2-cocycle term \eqref{KGB} satisfies the suitably generalized cocycle condition
\be
{K_{GB}}_{[(\xi_1,\omega_1),(\xi_2,\omega_2)],(\xi_3,\omega_3)}-\delta_3 {K_{P}}_{(\xi_1,\omega_1),(\xi_2,\omega_2)} + \text{cyclic}(1,2,3)=0.
\ee
Again, one needs to find another slicing to have this 2-cocycle term really at the center of the algebra. Nevertheless, the Gauss-Bonnet term wound not affect the balance equation \eqref{balance} for the same reason as the Pontryagin term.

Here, we present another way of separating the Gauss-Bonnet charge into an integrable part
\be
{\cal H}_{GB  (\xi,\omega)}^I=\frac{1}{8\pi G}\int_{\partial \Sigma} \td z \td \bz \bigg[\frac{\xbar\tau_0}{P}\delta_{\xi,\omega}\p_{\bz} \ln P + \frac{\tau_0}{P} \delta_{\xi,\omega}\p_z \ln P
+2 \frac{\rho_0}{P} \delta_{\xi,\omega} \frac{\mu_0}{P}\bigg],
\ee
and a flux part
\begin{multline}
{\cal F}_{GB (\xi,\omega)} (\delta \phi^\alpha;\phi^\alpha)=-\frac{1}{8\pi G}\int_{\partial \Sigma} \td z \td \bz \bigg[\frac{\xbar\tau_0}{P}\delta(\delta_{\xi,\omega}\p_{\bz} \ln P) + \frac{\tau_0}{P} \delta(\delta_{\xi,\omega}\p_z \ln P)\\
+2 \frac{\rho_0}{P} \delta (\delta_{\xi,\omega} \frac{\mu_0}{P})
+2\delta \frac{\mu_0}{P}\delta_{\xi,\omega}\frac{\rho_0}{P} + \delta_{\xi,\omega}\frac{\xbar\tau_0}{P} \delta\p_{\bz} \ln P+\delta_{\xi,\omega} \frac{\tau_0}{P}  \delta\p_z \ln P\bigg].
\end{multline}
The charge satisfies the modified bracket with no 2-cocycle term and the balance equation \eqref{balance} in this case can be directly verified. The Gauss-Bonnet charge of the Taub-NUT solution can be obtained as
\begin{multline}
{\cal H}_{GBT  (\xi,\omega)}^I=\frac{1}{8\pi G}\frac{iN}{r_+}\int_{\partial \Sigma} \td z \td \bz \bigg[\frac{  z}{(1+z\bz)^2}\left(\bY-\bz\p_{\bz} \bY-\frac12(1+z\bz) \p_z^2 Y\right)\\
- \frac{\bz}{(1+z\bz)^2}\left(Y - z\p_z Y -\frac12  (1+z\bz) \p_{\bz}^2 \bY\right)\bigg].
\end{multline}
All supertranslation charges of the Taub-NUT solution vanishes for the Gauss-Bonnet term. The zero mode of superrotation $(f=0,Y^z=Y^{\bz}=1)$ charge is
\be
J_{GB}=\frac{-i}{8\pi G}\frac{N}{r_+}\int_{\partial \Sigma} \td z \td \bz \left[ \frac{\bz-z}{(1 + z \bz)^2} \right].
\ee
Remarkably, we find that the zero mode of the Gauss-Bonnet superrotation charge and the zero mode of the Pontryagin superrotation charge \eqref{JA} are given in a very similar way. One is determined by the mass parameter while the other is determined by the NUT parameter. Considering the relation between the Pontryagin term and the Gauss-Bonnet term, we believe this is another evidence that the two parameters $M$ and $N$ are dual to each other.

\section{Conclusion and outlook}

We have shown that the surface charge from the Pontryagin term and Gauss-Bonnet term can also arise from a combination of $Y$ and $W$ ambiguity of the symplectic structure. The topological terms, such as Pontryagin term and Gauss-Bonnet term have significant effect on the 2-cocycle term of the near horizon charge algebra. The 2-cocycle term from the Pontryagin term and Gauss-Bonnet term are field dependent. In principle, such term may not be at the center of the algebra, i.e. commutes with all the charges. One possibility to improve the situation is to apply the change of slicing proposed in \cite{Adami:2020ugu,Ruzziconi:2020wrb,Adami:2021sko} to redefine the symmetry parameters. However, in the derivation of the Palatini charge \eqref{palatinicharge}, it is already assumed that the symmetry parameters are field independent \cite{Godazgar:2020kqd}. Hence the slicing is fixed and can not be changed in this work. We will stress this issue and fix the 2-cocycle term from Pontryagin term and Gauss-Bonnet term in a forthcoming work.

We use the Barnich-Troessaert prescription to define a modified charge algebra by splitting the charge into integrable and flux parts. The charges from the boundary terms only contribute to the flux part. In general, the integrable charge from Barnich-Troessaert prescription is different from the Noether charge. However it is conjectured in \cite{Freidel:2021cjp} and further verified in \cite{Adami:2022ktn} that the two charges can be made identical by introducing a $W$ term in the symplectic structure. We hope this $W$ term and our 2-cocycle term can play a role in resolving the Kerr/CFT and Wald entropy discrepancy in high derivative gravities
\cite{Krishnan:2009tj,Azeyanagi:2009wf,Liu:2021hvb}.

There could be a couple of applications of the near horizon charge from the topological terms. We point out some of them for future directions, for instance in null hypersurface thermodynamics \cite{Adami:2021kvx}, in the triangle relation \cite{Strominger:2017zoo} and in resolving the black hole information paradox \cite{Hawking:2016msc}.

\section*{Acknowledgments}

The authors thank Mahdi Godazgar, Hong L\"{u}, Roberto Oliveri, Shahin Sheikh-Jabbari, Jun-Bao Wu and Xiaoning Wu for discussions and comments. This work is supported in part by the National Natural Science Foundation of China under Grant No. 11905156, No. 12075166, No. 11675144, and No. 11935009.

\appendix

\section{Useful relations in Newman-Penrose formalism}
\label{relation}

The components of spin connection
\begin{align}
&\kappa=\Gamma_{311}=l^\nu m^\mu\nabla_\nu l_\mu,\;\;\pi=-\Gamma_{421}=-l^\nu \bar{m}^\mu\nabla_\nu n_\mu,\nn\\
&\epsilon=\half(\Gamma_{211}-\Gamma_{431})=\half(l^\nu n^\mu\nabla_\nu l_\mu - l^\nu \bar{m}^\mu\nabla_\nu m_\mu),\nn\\
&\nn\\
&\tau=\Gamma_{312}=n^\nu m^\mu\nabla_\nu l_\mu,\;\;\nu=-\Gamma_{422}=-n^\nu \bar{m}^\mu\nabla_\nu n_\mu,\nn\\
&\gamma=\half(\Gamma_{212}-\Gamma_{432})=\half(n^\nu n^\mu\nabla_\nu l_\mu - n^\nu \bar{m}^\mu\nabla_\nu m_\mu),\nn\\
&\nn\\
&\sigma=\Gamma_{313}=m^\nu m^\mu\nabla_\nu l_\mu,\;\;\mu=-\Gamma_{423}=-m^\nu \bar{m}^\mu\nabla_\nu n_\mu,\nn\\
&\beta=\half(\Gamma_{213}-\Gamma_{433})=\half(m^\nu n^\mu\nabla_\nu l_\mu - m^\nu \bar{m}^\mu\nabla_\nu m_\mu),\nn\\
&\nn\\
&\rho=\Gamma_{314}=\bar{m}^\nu m^\mu\nabla_\nu l_\mu,\;\;\lambda=-\Gamma_{424}=-\bar{m}^\nu \bar{m}^\mu\nabla_\nu n_\mu,\nn\\
&\alpha=\half(\Gamma_{214}-\Gamma_{434})=\half(\bar{m}^\nu n^\mu\nabla_\nu l_\mu - \bar{m}^\nu \bar{m}^\mu\nabla_\nu m_\mu).\nn
\end{align}
Ten components of the Weyl tensor
\begin{align}
\Psi_0=-C_{1313},\;\;\Psi_1=-C_{1213},\;\;\Psi_2=-C_{1342},\;\;\Psi_3=-C_{1242},\;\;\Psi_4=-C_{2424}.\nn
\end{align}
Four real components of the Ricci tensor
\be
\begin{split}
\Phi_{00}=-\half R_{11},&\;\;\Phi_{22}=-\half R_{22},\\
\Phi_{11}=-\dfrac{1}{4}(R_{12}+R_{34}),&\;\;\Lambda=\dfrac{1}{24}R=\dfrac{1}{12} (R_{12}-R_{34}),
\end{split}\nn
\ee
where $\Lambda$ is the cosmological constant and three complex ones
\begin{align}
&\Phi_{02}=-\half R_{33},\;\;\Phi_{20}=-\half R_{44},\nn\\
&\Phi_{01}=-\half R_{13},\;\;\Phi_{10}=-\half R_{14},\nn\\
&\Phi_{12}=-\half R_{23},\;\;\Phi_{21}=-\half R_{24}.\nn
\end{align}

The orthogonality conditions and normalization conditions \eqref{tetradcondition} of the basis vectors yield the following relations
\be
\begin{split}\label{1}
&l^\nu \n_\nu l_\mu=(\epsilon+\bar \epsilon)l_\mu -\kappa \bm_\mu - \bar \kappa m_\mu,\\
&n^\nu \n_\nu l_\mu=(\gamma+\bar \gamma)l_\mu -\tau \bm_\mu - \bar \tau m_\mu,\\
&m^\nu \n_\nu l_\mu=(\beta+\bar \alpha)l_\mu -\sigma \bm_\mu - \bar \rho m_\mu,\\
&\bm^\nu \n_\nu l_\mu=(\alpha + \bar \beta)l_\mu - \rho \bm_\mu -\bar\sigma m_\mu,
\end{split}
\ee
\be
\begin{split}\label{2}
&l^\nu \n_\nu n_\mu=-(\epsilon+\bar \epsilon)n_\mu + \bar\pi \bm_\mu + \pi m_\mu,\\
&n^\nu \n_\nu n_\mu=-(\gamma+\bar \gamma)n_\mu + \bar\nu \bm_\mu + \nu m_\mu,\\
&m^\nu \n_\nu n_\mu=-(\beta+\bar \alpha)n_\mu + \bar\lambda \bm_\mu + \mu m_\mu,\\
&\bm^\nu \n_\nu n_\mu=-(\alpha + \bar \beta)n_\mu + \bar\mu \bm_\mu + \lambda m_\mu,\\
\end{split}
\ee
\be
\begin{split}\label{3}
&l^\nu \n_\nu m_\mu=(\epsilon - \bar\epsilon) m_\mu - \kappa n_\mu + \bar\pi l_\mu,\\
&n^\nu \n_\nu m_\mu=(\gamma - \bar\gamma) m_\mu  - \tau n_\mu + \bar\nu l_\mu,\\
&m^\nu \n_\nu m_\mu= (\beta - \bar\alpha) m_\mu - \sigma n_\mu + \bar\lambda l_\mu,\\
&\bm^\nu \n_\nu m_\mu= (\alpha - \bar\beta) m_\mu - \rho n_\mu + \bar\mu l_\mu,
\end{split}
\ee
After setting $\epsilon=\kappa=\pi=0$, the geodesic deviation of $l$ is
\begin{multline}\label{Bl}
B_{\nu\mu}^l=\n_\nu l_\mu=(\gamma+\bar \gamma)l_\mu l_\nu - \tau \bm_\mu l_\nu - \bar \tau m_\mu l_\nu - (\beta+\bar \alpha)l_\mu \bm_\nu - (\bar \beta+\alpha)l_\mu m_\nu \\ + \sigma \bm_\mu \bm_\nu + \bar \rho m_\mu \bm_\nu + \bar\sigma m_\mu m_\nu + \rho \bm_\mu m_\nu.
\end{multline}
The transverse part is
\be\label{Bl0}
\hat{B}_{\nu\mu}^l=\sigma \bm_\mu \bm_\nu + \bar \rho m_\mu \bm_\nu + \bar\sigma m_\mu m_\nu + \rho \bm_\mu m_\nu.
\ee
It is clear that $\rho=\bar\rho$ means $l$ is hypersurface orthogonal and further setting $\tau=\bar\alpha+\beta$ leads to the fact that $\n_\nu l_\mu$ is a symmetric tensor, then $l$ is the gradient of a scalar field.

On the horizon $r=0$, $n$ is tangent to geodesic. The geodesic deviation of $n$ is
\begin{multline}\label{Bn}
B_{\nu\mu}^n=\n_\nu n_\mu=-(\gamma+\bar\gamma)n_\mu l_\nu + \bar\nu \bm_\mu l_\nu + \nu m_\mu l_\nu + (\bar\alpha + \beta ) n_\mu \bm_\nu+ (\alpha + \bar\beta ) n_\mu m_\nu\\ - \bar\lambda \bm_\mu \bm_\nu - \lambda m_\mu m_\nu - \mu m_\mu \bm_\nu - \bar \mu \bm_\mu m_\nu.
\end{multline}
The transverse part is
\be\label{Bn0}
\hat{B}_{\nu\mu}^n= - \bar\lambda \bm_\mu \bm_\nu - \lambda m_\mu m_\nu - \mu m_\mu \bm_\nu - \bar \mu \bm_\mu m_\nu.
\ee

\section{NP equations}
\label{NP}
Considered as directional derivatives, the basis vectors are assigned with special symbols
\begin{align}
D=l^\mu\p_\mu,\;\;\;\;\Delta=n^\mu\p_\mu,\;\;\;\;\delta=m^\mu\p_\mu.
\end{align}
\textbf{Radial equations}
\bea
&&D\rho =\rho^2+\sigma\xbar\sigma,\label{R1}\\
&&D\sigma=2\rho \sigma + \Psi_{0},\label{R2}\\
&&D\tau =\tau \rho +  \xbar \tau \sigma   + \Psi_1 ,\label{R3}\\
&&D\alpha=\rho  \alpha + \beta \xbar \sigma  ,\label{R4}\\
&&D\beta  =\alpha \sigma + \rho  \beta + \Psi_{1},\label{R5}\\
&&D\gamma=\tau \alpha +  \xbar \tau \beta  + \Psi_2,\label{R6}\\
&&D\lambda=\rho\lambda + \xbar\sigma\mu ,\label{R7}\\
&&D\mu =\rho \mu + \sigma\lambda + \Psi_{2},\label{R8}\\
&&D\nu =\xbar\tau \mu + \tau  \lambda + \Psi_3,\label{R9}\\
&&DU=\xbar\tau\omega+\tau\xbar\omega - (\gamma+\xbar\gamma),\label{R10}\\
&&DX^A=\xbar\tau L^A + \tau\bar L^A,\label{R11}\\
&&D\omega=\rho\omega+\sigma\xbar\omega-\tau,\label{R12}\\
&&DL^A=\rho L^A + \sigma \bar L^A,\label{R13}\\
&&D\Psi_1 - \xbar\delta \Psi_0 =  4 \rho \Psi_1 - 4\alpha \Psi_0,\label{R14}\\
&&D\Psi_2 - \xbar\delta \Psi_1 =   3\rho \Psi_2  - 2 \alpha \Psi_1- \lambda \Psi_0,\label{R15}\\
&&D\Psi_3 - \xbar\delta \Psi_2 =  2\rho \Psi_3 - 2\lambda \Psi_1,\label{R16}\\
&&D\Psi_4 - \xbar\delta \Psi_3 = \rho  \Psi_4 + 2 \alpha \Psi_3 - 3 \lambda \Psi_2,\label{R17}
\eea

\textbf{Non-radial equations}
\bea
&&\Delta\lambda  = \xbar\delta\nu- (\mu + \xbar\mu)\lambda - (3\gamma - \xbar\gamma)\lambda + 2\alpha \nu - \Psi_4,\label{H1}\\
&&\Delta\rho= \xbar\delta\tau- \rho\xbar\mu - \sigma\lambda  -2\alpha \tau + (\gamma + \xbar\gamma)\rho  - \Psi_2,\label{H2}\\
&&\Delta\alpha = \xbar\delta\gamma +\rho \nu - (\tau + \beta)\lambda + (\xbar\gamma - \gamma -\xbar \mu)\alpha  -\Psi_3 ,\label{H3}\\
&&\Delta \mu=\delta\nu-\mu^2 - \lambda\xbar\lambda - (\gamma + \xbar\gamma)\mu   + 2 \beta \nu ,\label{H4}\\
&&\Delta \beta=\delta\gamma - \mu\tau + \sigma\nu + \beta(\gamma - \xbar\gamma -\mu) - \alpha\xbar\lambda ,\label{H5}\\
&&\Delta \sigma=\delta\tau - \sigma\mu - \rho\xbar\lambda - 2 \beta \tau + (3\gamma - \xbar\gamma)\sigma  ,\label{H6}\\
&&\Delta \omega=\delta U +\xbar\nu -\xbar\lambda\xbar\omega + (\gamma-\xbar\gamma-\mu)\omega,\label{H7}\\
&&\Delta L^A=\delta X^A - \xbar\lambda \bar L^A + (\gamma-\xbar\gamma-\mu)L^A,\label{H8}\\
&&\delta\rho - \xbar\delta\sigma=\rho\tau - \sigma (3\alpha - \xbar\beta)   - \Psi_1 ,\label{H9}\\
&&\delta\alpha - \xbar\delta\beta=\mu\rho - \lambda\sigma + \alpha\xbar\alpha + \beta\xbar\beta - 2 \alpha\beta - \Psi_2,\label{H10}\\
&&\delta\lambda - \xbar\delta\mu= \mu \xbar\tau + \lambda (\xbar\alpha - 3\beta) - \Psi_3 ,\label{H11}\\
&&\delta \xbar\omega-\bar\delta\omega=\mu - \xbar\mu - (\alpha - \xbar\beta) \omega +  (\xbar\alpha - \beta)\xbar\omega,\label{H12}\\
&&\delta \bar L^A - \bar\delta L^A= (\xbar\alpha - \beta)\bar L^A -  (\alpha - \xbar\beta) L^A ,\label{H13}\\
&&\Delta\Psi_0 - \delta \Psi_1 = (4\gamma -\mu)\Psi_0 - (4\tau + 2\beta)\Psi_1 + 3\sigma \Psi_2,\label{H14}\\
&&\Delta\Psi_1 - \delta \Psi_2 = \nu\Psi_0 + (2\gamma - 2\mu)\Psi_1 - 3\tau \Psi_2 + 2\sigma \Psi_3 ,\label{H15}\\
&&\Delta\Psi_2 - \delta \Psi_3 = 2\nu \Psi_1 - 3\mu \Psi_2 + (2\beta - 2\tau) \Psi_3 + \sigma \Psi_4,\label{H16}\\
&&\Delta\Psi_3 - \delta \Psi_4 = 3\nu \Psi_2 - (2\gamma + 4\mu) \Psi_3 + (4\beta - \tau) \Psi_4,\label{H17}
\eea

\section{Details for deriving the solution space}
\label{NU solution}

The radial equations are organized in different groups. The first group is \eqref{R1} and \eqref{R2}. Once the whole series of $\Psi_0$ is given as initial data as \eqref{psi0}, $\rho$ and $\sigma$ are solved out as \eqref{rho} and \eqref{sigma}. Inserting the solutions of $\rho$ and $\sigma$ into \eqref{R13}, one gets $L^A$ as \eqref{Lz} and \eqref{Lzb}, then $L_A$ is derived by the condition $L_AL^A=0,\;L_A\bar L^A=-1$. The second group of radial equations consists of \eqref{R4}, \eqref{R5}, \eqref{R12} and \eqref{R14}. One can work out $\alpha$, $\beta$, $\omega$ and $\Psi_1$ as \eqref{alpha}, \eqref{beta}, \eqref{omega}, and \eqref{psi1} respectively, then $\tau$ from gauge condition $\tau=\xbar\alpha+\beta$. Inserting $\tau$ and $L^A$ into \eqref{R11}, $X^A$ could be obtained as \eqref{XA}. The third group of radial equations include \eqref{R7}, \eqref{R8}, and \eqref{R15}. One can just apply the same method as the first two groups to solve out $\mu$, $\lambda$, and $\Psi_2$, which are given in \eqref{mu}, \eqref{lambda}, and \eqref{psi2}. Then, $\gamma$ is derived from \eqref{R6} as \eqref{gamma},  $U$ is derived from \eqref{R10} as \eqref{U},  $\Psi_3$ is derived from \eqref{R16} as \eqref{psi3},  $\nu$ is derived from \eqref{R9} as \eqref{nu}, and finally $\Psi_4$ is derived from \eqref{R17} as \eqref{psi4}.

Inserting the solutions of the radial equations into the non-radial equations, more constraints are obtained for the integration constants:

\eqref{H13} yields \eqref{alpha0} and \eqref{beta0}.

\eqref{H8} yields \eqref{mu0}.

\eqref{H4} yields \eqref{gamma0}.

\eqref{H1} yields \eqref{psi40}.

 \eqref{H3} yields \eqref{psi30}.

\eqref{H2} yields \eqref{psi20}.

\eqref{H9} yields \eqref{psi10}.

\eqref{H11} yields \eqref{tau0}.

\eqref{H10} yields \eqref{rho0}.

\eqref{H6} yields \eqref{sigma0}.

\eqref{H14} yields \eqref{psi00}.

The rest non-radial equations \eqref{H5}, \eqref{H7}, \eqref{H8}, \eqref{H12}, and \eqref{H15}-\eqref{H17} are satisfied automatically.

\section{Details for deriving the asymptotic symmetry}
\label{asg}

The gauge conditions yield
\begin{itemize}
\item
  $0=\delta_{\xi,\omega}\; e_1^u=-\p_r \xi^u \Longrightarrow
  \xi^u=f(u,z,\bz)$.
\item
  $0=\delta_{\xi,\omega}\; e_2^u=-e_2^\alpha \p_\alpha f + \omega^{12}
  \Longrightarrow \omega^{12}=\p_u f + X^A \p_A f$.
\item
  $0=\delta_{\xi,\omega}\; e_3^u=-e_3^\alpha \p_\alpha f + \omega^{24}
  \Longrightarrow \omega^{24}= L^A \p_A f$.
\item
  $0=\delta_{\xi,\omega}\; e_4^u=-e_4^\alpha \p_\alpha f + \omega^{23}
  \Longrightarrow \omega^{23}= \bL^A \p_A f$.
\item
  $0=\delta_{\xi,\omega}\; e_1^r=-e_1^\alpha \p_\alpha \xi^r +
  \omega^{2a}e_a^r \Longrightarrow \xi^r=-\p_u f r + Z(u,z,\bz) + \p_A
  f \int^r_0 \td r[\omega \bL^A + \bomega L^A - X^A]$.
\item
  $0=\delta_{\xi,\omega}\; e_1^A=-e_1^\alpha \p_\alpha \xi^A +
  \omega^{2a}e_a^A \Longrightarrow \xi^A=Y^A(u,z,\bz) + \p_B f
  \int^r_0 \td r[L^A \bL^B + \bL^AL^B ]$.
\item
  $\delta_{\xi,\omega}\;\bar\pi=0\iff 0=\delta_{\xi,\omega}\;
  \Gamma_{321}=l^\mu \p_\mu \omega^{41} + \Gamma_{32a} \omega^{2a}
  \Longrightarrow \omega^{14}=\omega^{14}_0(u,z,\bz) - \p_A f
  \int^r_0 \td r[\bar{\lambda} \bL^A + \bar{\mu} L^A]$.
\item $\delta_{\xi,\omega}\;\pi=0\iff 0=\delta_{\xi,\omega}\;
  \Gamma_{421}=l^\mu \p_\mu \omega^{31} + \Gamma_{42a} \omega^{2a}
  \Longrightarrow
\omega^{13}=\omega^{13}_0(u,z,\bz) - \p_A f \int^r_0
\td r[\lambda L^A + \mu \bL^A]$.
\item
  $\delta_{\xi,\omega}\;(\epsilon-\bar\epsilon)=0\iff
  0=\delta_{\xi,\omega}\; \Gamma_{431}=l^\mu \p_\mu \omega^{43} +
  \Gamma_{43a} \omega^{2a} \Longrightarrow
  \omega^{34}=\omega^{34}_0(u,z,\bz) + \p_A f \int^r_0
  \td r[(\bar{\alpha}-\beta) \bL^A - (\alpha - \bar{\beta}) L^A]$.

\item  Direct computation shows that the conditions  $\epsilon+\bar\epsilon=0=\kappa=\bar\kappa$, $\rho-\bar\rho=0$ and
  $\tau-\bar\alpha-\beta=0$ do not lead to new constraints on the symmetry parameters. These conditions hold as a consequence of the tetrad conditions imposed in \eqref{tetrad} when the NP equations are satisfied.
\end{itemize}
Before checking the constraints from the boundary conditions, we set $Z=0$. The reason is that we want to fix the $r=0$ null hypersurface to be the boundary. In the near horizon expansion, we will always assume that the horizon is located at the zero value of the radial coordinate.  A shift in $r$ direction, i.e., $\tilde{r}=r+Z$ combined with transformations in other coordinates can set $\tilde{r}=0$ hypersurface to be null but will also change the boundary to be the $\tilde{r}=0$ hypersurface. Somehow this can be understood as the fact that the existence of a boundary at $r=0$ breaks the translational invariance along $r$ direction, see also in \cite{Donnay:2015abr,Adami:2021nnf}.

The fall-off conditions yield
\begin{itemize}
\item $\delta_{\xi,\omega}\; e_3^r=O(r) \Longrightarrow \omega^{41}_0=0$.
\item $\delta_{\xi,\omega}\; e_4^r=O(r) \Longrightarrow \omega^{31}_0=0$.
\item $\delta_{\xi,\omega}\; e_2^A=O(r) \Longrightarrow \p_u Y^A=0$.
\item $\delta_{\xi,\omega}\; e_3^z=O(r) \Longrightarrow \p_{\bz} Y^z =0$.
\item $\delta_{\xi,\omega}\; e_4^{\bz}=O(r) \Longrightarrow \p_z Y^{\bz} =0$.
\item $\delta_{\xi,\omega}\;\text{Im}[ e_3^{\bz}]=O(r) \Longrightarrow \omega^{43}_0=\frac12(\p_z Y^z - \p_{\bz} Y^{\bz})$.
\end{itemize}
The boundary conditions on $e_2^r$, $\lambda$, $\nu$ and $\text{Im}[\mu]$ do not lead to new constraint on the symmetry parameters as confirmed by direct computation.

\section{Near horizon expansion for Taub-NUT solution}
\label{TN}

Here, we transform the near horizon expansions of the Taub-NUT solution \cite{Taub:1950ez,Newman:1963yy} into the gauge we used in this work. The procedure is similar to the case of the asymptotic expansions in the Bondi gauge in \cite{Godazgar:2019dkh,Godazgar:2019ikr}. The Taub-NUT solution in the complex stereographic coordinates adapted to the NP conventions is \cite{Griffiths:2009dfa}
\be\label{Taub-NUT-metric}
\td s^2 = f(\chi)\left(\td t + 2 i N \frac{\zeta \td \bar\zeta - \bar\zeta \td \zeta}{1+\zeta\bar\zeta}\right)^2 - \frac{\td \chi^2}{f(\chi)} - (\chi^2+N^2)\frac{4\td \zeta \td \bar\zeta}{(1+\zeta\bar\zeta)^2},
\ee
where $f(\chi)=\frac{\chi^2 - 2 M\chi - N^2}{\chi^2 + N^2}$. $N$ is the NUT parameter and $M$ is the mass parameter. We perform the following coordinates transformation
\be
\begin{split}
&t=u-\frac{2N^2z\bz}{r_+^2 + N^2}r+\frac{N^2z\bz\left[2N^2 r_+ + 2 r_+^3 + N^2z\bz(r_+-r_-)\right]}{(r_+^2 + N^2)^3}r^2+O(r^3),\\
&\chi=r_+ + r-\frac{N^2(r_+-r_-)}{(r_+^2 + N^2)^2}r^2+O(r^3),\\
&\zeta=z+\frac{iNz(1+z\bz)}{r_+^2 + N^2}r-\frac{iNr_+z(1+z\bz)+N^2z(1+z\bz)^2}{(r_+^2 + N^2)^2}r^2+O(r^3),
\end{split}
\ee
where $r_{\pm}=M\pm\sqrt{M^2+N^2}$. The line-element in the near horizon expansion is
\begin{multline}
\td s^2 = \frac{r}{r_+}\td u^2 + 2\td u \td r - \frac{2 i N \bz}{r_+(1+z\bz)}r \td u \td z + \frac{2 i N z}{r_+(1+z\bz)}r \td u \td \bz\\
+\frac{4N\left[i N^2 - N(r_+ - r_-) + i r_+^2\right]\bz^2}{(r_+^2 + N^2)(1+z\bz)^2}r\td z^2  -\frac{4N\left[i N^2 + N(r_+ - r_-) + i r_+^2\right]z^2}{(r_+^2 + N^2)(1+z\bz)^2}r\td \bz^2 \\
-\left\{\frac{4(r_+^2 + N^2)}{(1+z\bz)^2}+ \frac{8\left[r_+^3 + N^2\left(r_+-z\bz(r_+-r_-)\right)\right]}{(r_+^2 + N^2)(1+z\bz)^2}r\right\}\td z\td\bz+O(r^2).
\end{multline}
We have computed more orders but it is not relevant to the charge we are dealing with. Comparing to our near horizon line-element
\begin{multline}
\td s^2 = 4\gamma_0 r\td u^2+ 2\td u \td r + \frac{4\xbar\tau_0}{P} r\td u \td z + \frac{4 \tau_0}{P} r \td u \td \bz\\
+\frac{2\xbar\sigma_0}{P^2}r\td z^2  + \frac{2\sigma_0}{P^2}r\td \bz^2 -(\frac{1}{P^2} - \frac{2\rho_0}{P^2}r) \td z\td\bz+O(r^2),
\end{multline}
one can find, for the Taub-NUT solution, that
\be\label{taubnutNP}
\begin{split}
&\frac{1}{P^2}=\frac{4(r_+^2 + N^2)}{(1+z\bz)^2},\quad \gamma_0=  \frac{1}{4r_+},\quad \frac{\tau_0}{P}= \frac{i N  z}{2r_+(1+z\bz)},\\
&\mu_0=0,\quad \sigma_0 = - \frac{N(ir_+ + N) z^2}{2(r_+-r_-)r_+^2},\quad \rho_0=-\frac{\left[r_+^3 + N^2\left(r_+-(r_+-r_-)z\bz\right)\right]}{(r_+^2 + N^2)^2}.
\end{split}
\ee

\bibliography{ref}

\end{document}